\documentclass[10pt]{article}
\pdfoutput=1
\usepackage{threeparttable,booktabs,graphicx}
\usepackage{cite}
\usepackage[auth-lg]{authblk}
\usepackage[margin=2cm]{geometry}

\date{}

\title{\bf \Large Structural properties of thin-film ferromagnetic topological insulators}

\author[1]{C.~L.~Richardson}
\author[1]{J.~M.~Devine-Stoneman}
\author[1]{G.~Divitini}
\author[1]{M.~E.~Vickers}
\author[2,3]{C.-Z.~Chang}
\author[1]{M.~Amado}
\author[2]{J.~S.~Moodera}
\author[1,*]{J.~W.~A.~Robinson}
\affil[1]{\footnotesize University of Cambridge, Department of Materials Science and Metallurgy, Cambridge, CB3 0FS, UK}
\affil[2]{\footnotesize Massachusetts Institute of Technology, Francis Bitter National Magnet Laboratory, Cambridge, MA 02139, USA}
\affil[3]{\footnotesize Pennsylvania State University, Department of Physics, State College, PA 16802-6300, USA}
\affil[*]{\footnotesize jjr33@cam.ac.uk}

\begin{document} \flushbottom
\maketitle
\thispagestyle{empty}

\begin{abstract}
We present a comprehensive study of the crystal structure of the thin-film, ferromagnetic topological insulator (Bi,~Sb)$_{2-x}$V$_x$Te$_3$. The dissipationless quantum anomalous Hall edge states it manifests are of particular interest for spintronics, as a natural spin filter or pure spin source, and as qubits for topological quantum computing.
For ranges typically used in experiments, we investigate the effect of doping, substrate choice and film thickness on the (Bi,~Sb)$_2$Te$_3$ unit cell using high-resolution X-ray diffractometry. Scanning transmission electron microscopy and energy-dispersive X-ray spectroscopy measurements provide local structural and interfacial information.
We find that the unit cell is unaffected in-plane by vanadium doping changes, and remains unchanged over a thickness range of 4--10 quintuple layers (1~QL $\approx$ 1~nm). The in-plane lattice parameter (\emph{a}) also remains the same in films grown on different substrate materials. However, out-of-plane the \emph{c}-axis is reduced in films grown on less closely lattice-matched substrates, and increases with the doping level.
\end{abstract}

\section*{Introduction}
The quantum anomalous Hall effect (QAHE) allows the resistance quantization and dissipationless edge states seen in the quantum Hall effect \cite{vonKlitzing1980}, but without the need for an applied magnetic field. The topologically-protected edge states of the QAHE are also chiral and spin-polarised \cite{Bernevig2006,Yu2010}, acting as a natural spin filter. Additional applications are starting to be explored, for example in spintronics as a pure spin current source or detector \cite{Gotte2016}, and in topological quantum computing \cite{HasanKane2010,Qi2010}.

In experimental devices, the QAHE is remarkably robust, and has been observed in chromium- and vanadium-doped (Bi,~Sb)$_2$Te$_3$ across a range of film thicknesses, and grown on multiple substrates \cite{Chang2013,Checkelsky2014,Grauer2015,Kou2014,Bestwick2015}. Quantization has been improved by reducing film thickness and doping with vanadium to suppress dissipative channels \cite{Chang2015a}, but whether these variables affect the crystal structure and electronic band structure has largely been overlooked experimentally. To date, the \emph{a}-axis parameter has only been determined for Cr-doped (Bi,~Sb)$_2$Te$_3$ grown on SrTiO$_3$~(1~1~1) with or without a Te capping layer \cite{Park2015}, with no systematic study of the role of thickness or doping level. This is despite the potential for uniaxial or biaxial strain in the films to either drive these materials into the topologically trivial (non-QAH) regime by altering the band structure \cite{Aramberri2016}, or affect the fabrication and performance of ferromagnetic/non-ferromagnetic topological insulator heterostructure devices due to lattice mismatch \cite{Yasuda2016,He2016,Mogi2017}.

Here, we conduct a comprehensive study of the effect of thickness, vanadium-doping and substrate choice on the crystal structure of MBE-grown (Bi,~Sb)$_{2-x}$V$_x$Te$_3$ thin films, and compare to existing results in the literature. We use high-resolution X-ray diffractometry (HRXRD) to determine the in- and out-of-plane lattice parameters, orientation relationships and epitaxial quality of the film, substrate, and Te capping layer. The HRXRD data is supported by information about the local structure and doping, provided by scanning transmission electron microscopy (STEM) and energy-dispersive X-ray spectroscopy (EDX).

\section*{Results}
\subsection*{Crystallinity and elemental composition}
\begin{figure}[ht]
\centering
\includegraphics[width=12.75cm]{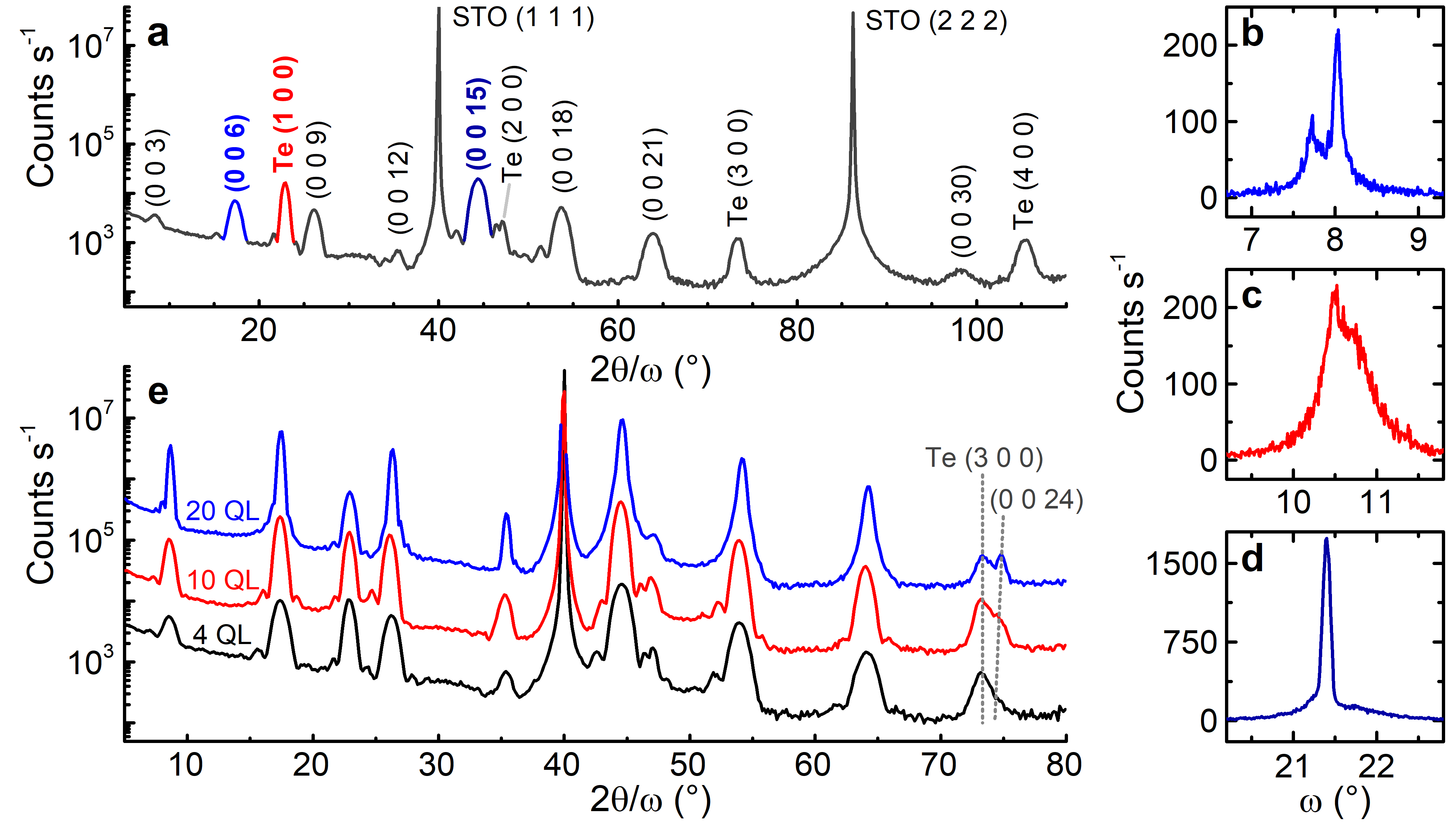}
\caption{High-resolution X-ray diffractometry of vanadium-doped (Bi,~Sb)$_2$Te$_3$ thin films on SrTiO$_3$~(1~1~1). \textbf{a} Indexed 2$\theta$/$\omega$ scan of a 4 QL film (quintuple layer; 1 QL $\approx$ 1 nm) with 10~nm tellurium capping layer. (Bi,~Sb)$_{2-x}$V$_x$Te$_3$ peaks are unnamed. \textbf{b-d} Rocking curves of (0 0 6), Te (1 0 0) and (0 0 15) peaks (shown in \textbf{a} in blue, red and dark blue, respectively). \textbf{e} 2$\theta$/$\omega$ scans of 4 (black), 10 (red) and 20 QL (blue) films (10 and 20 QL data offset for clarity). Grey dashed lines are guides to the eye, showing the contraction of the lattice as thickness increases.}
\label{epitaxy}
\end{figure}

We first demonstrate the quality of the films with symmetrical high-resolution X-ray diffraction measurements, probing planes parallel to the substrate. Figure \ref{epitaxy}a shows a 2$\theta$/$\omega$ scan of a 4~QL (quintuple layer, each QL~$\approx$1~nm) film of (Bi,~Sb)$_{1.89}$V$_{0.11}$Te$_3$ on a SrTiO$_3$~(1~1~1) substrate, with a 10~nm Te capping layer. We observe (Bi,~Sb)$_{2-x}$V$_{x}$Te$_3$ peaks at (0 0 3$n$), as expected for its space group $R\bar{3}m$, and tellurium peaks at ($m$ 0 0), echoing previous results on Cr-doped (Bi,~Sb)$_2$Te$_3$ \cite{Chang2015a,Park2015}.

Both layers grow epitaxially, as shown by the rocking curves in Figure \ref{epitaxy}b-\ref{epitaxy}d (rocking curves taken on the (Bi,~Sb)$_{2-x}$V$_{x}$Te$_3$ (0~0~6), Te (1 0 0) and (Bi,~Sb)$_{2-x}$V$_{x}$Te$_3$ (0 0 15) peaks, respectively). The (Bi,~Sb)$_{1.89}$V$_{0.11}$Te$_3$ (Figure \ref{epitaxy}b and \ref{epitaxy}d) has a smaller full-width at half-maximum (FWHM, Gaussian fit) than tellurium: $\sim$0.11$^{\circ}$ as opposed to 0.65$^{\circ}$. The (Bi,~Sb)$_{2-x}$V$_x$Te$_3$ (0~0~6) curve shows dual peaks $\sim$0.15$^{\circ}$ from the expected value of $\omega$, consistent with the observation of twinned crystal domains in Cr-doped films \cite{Richardella2015}.

These results hold for all samples grown on SrTiO$_3$~(1~1~1), even with changes to doping and thickness. Figure \ref{epitaxy}e shows 2$\theta$/$\omega$ scans of samples with $x$~=~0.06-0.07 and thicknesses of 4, 10 and 20~QLs (black, red and blue, respectively). The plots are offset for clarity. The changes in doping and thickness do not affect the relative intensities of the peaks. The (Bi, Sb)$_{2-x}$V$_{x}$Te$_3$ \emph{c}-axis appears to shorten as the thickness increases; this is clearly seen in the 2$\theta$ position of the (0 0 24) peak as it changes relative to the Te~(3 0 0) peak, which in turn stays constant with respect to SrTiO$_3$ (1 1 1). However, averages of Gaussian fits to 3 reflections show that 4 and 10~QL films have approximately equal thickness (30.51 $\pm$ 0.05 \AA and 30.54 $\pm$ 0.04 \AA respectively, compared to \emph{c}~=~30.44 $\pm$ 0.02 \AA for the 20~QL film).

HRXRD measurements of 10~QL films grown on Al$_2$O$_3$~(0~0~0~1) and Si~(1~1~1) show very similar epitaxial growth and out-of-plane lattice parameters (See Supplementary Figure S1). The (Bi,~Sb)$_{2-x}$V$_{x}$Te$_3$ lattice constant \emph{c}~=~30.47~$\pm$~0.08~{\AA} and 30.45~$\pm$~0.07~\AA, respectively, a slight decrease compared to 30.54~$\pm$~0.03~{\AA} on SrTiO$_3$~(1~1~1) (10~QL film). On an Al$_2$O$_3$~(0~0~0~1) substrate, rocking curves on (Bi,~Sb)$_{2-x}$V$_{x}$Te$_3$ (0 0 15) and Te~(1~0~0) have a FWHM of 0.12$^{\circ}$ and 0.77$^{\circ}$, respectively, similar to the SrTiO$_3$~(1~1~1) samples. Growth appears to be more disordered on Si~(1~1~1), where the (Bi, Sb)$_{2-x}$V$_{x}$Te$_3$ (0 0 15) rocking curve has a larger FWHM of $\sim$0.36$^{\circ}$.

\begin{figure}[ht]
\centering
\includegraphics[width=12.75cm]{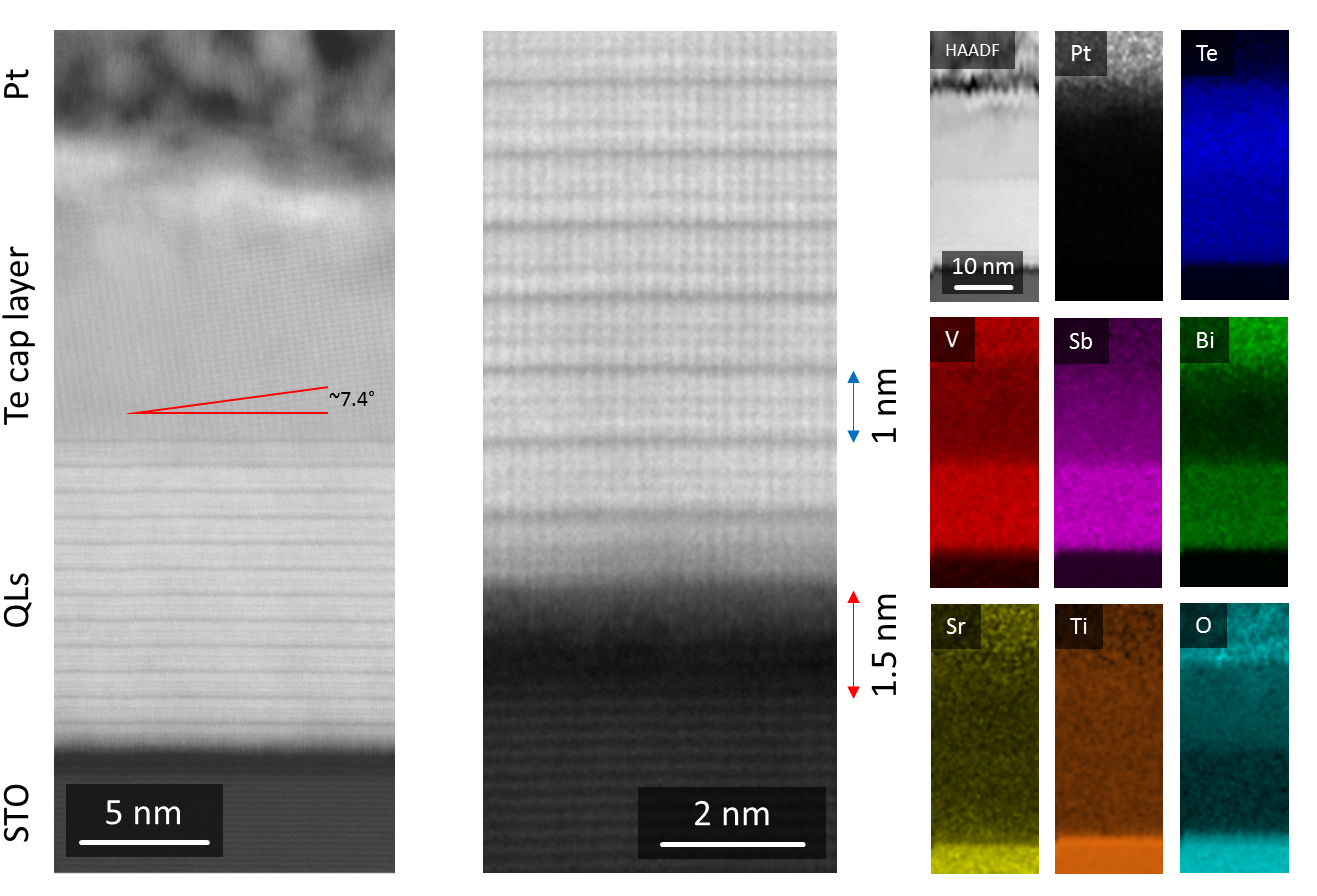}
\caption{Scanning transmission electron microscopy (STEM) and elemental analysis of (Bi,~Sb)$_{2-x}$V$_{x}$Te$_3$ on SrTiO$_3$. \textbf{a} STEM cross-sectional view of the film. \textbf{b} HRSTEM view of the substrate-film interface. \textbf{c} HAADF reference image and EDX elemental maps.}
\label{TEM}
\end{figure}

We now investigate the local crystallinity and interfaces using scanning transmission electron microscopy (STEM). The High Angle Annular Dark Field (HAADF) signal, generating the images shown in Figure~\ref{TEM}, is proportional to the local thickness and atomic number. For a relatively homogeneous thickness, such as a TEM lamella, the brightness is proportional to the average Z for a given pixel such that brighter atomic columns correspond to heavier atoms.

\begin{figure}[ht]
\centering
\includegraphics[width=12.75cm]{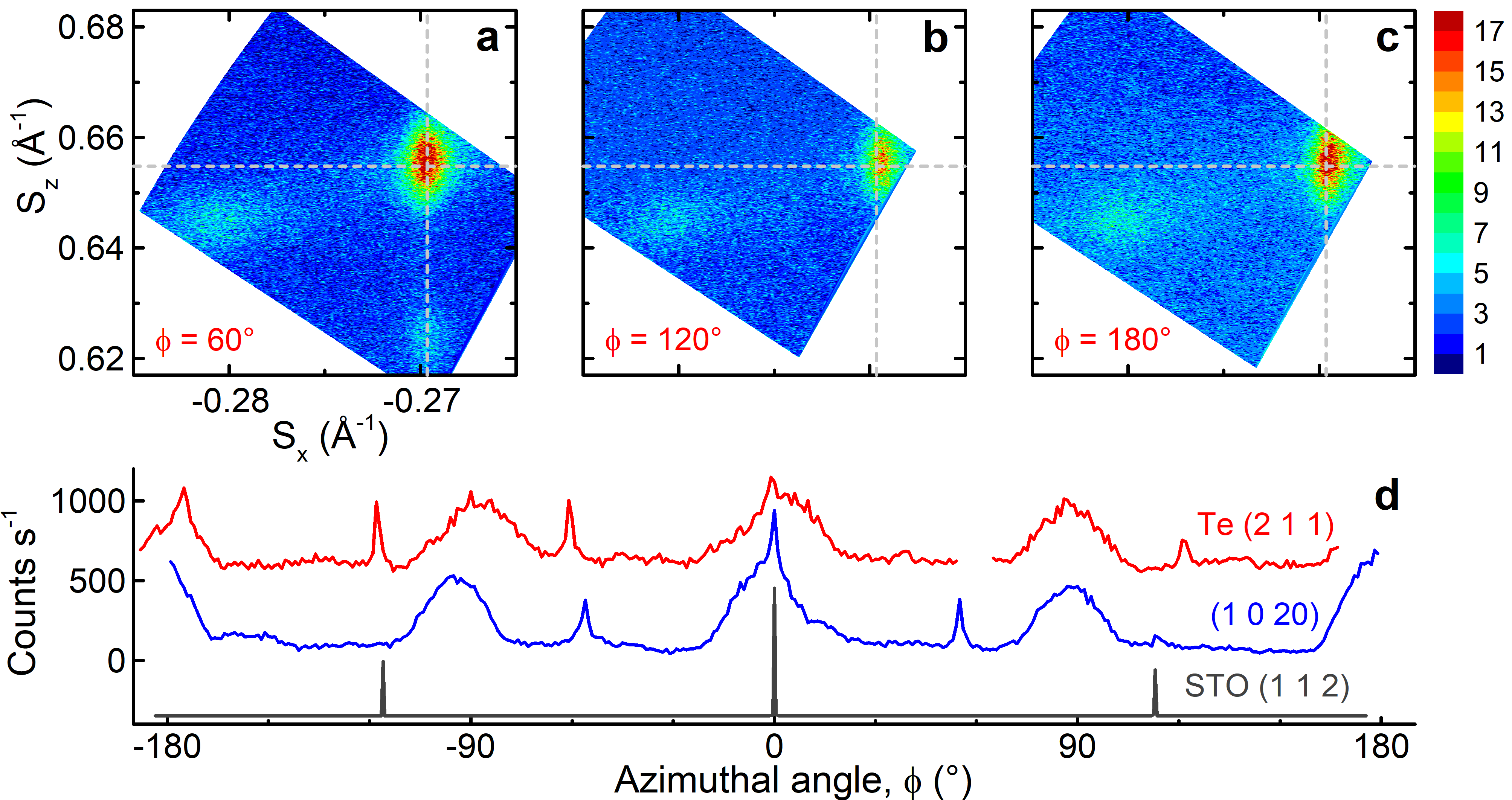}
\caption{In-plane orientation of layers. \textbf{a-c} Reciprocal space maps of the (1 0 20) and Te (2 1 1) peaks of a 10 QL, x~=~0.07 film, at azimuthal angles $\phi$~=~60$^{\circ}$, 120$^{\circ}$ and 180$^{\circ}$ respectively. Grey dashed lines indicate fitted peak value from data in Figure \ref{thickness}\textbf{b}. Film peaks measured with respect to SrTiO$_3$~(1~1~2). \textbf{d} Azimuthal angle scans of substrate (grey, STO (1 1 2): 2$\theta$~=~57.84$^{\circ}$, offset~=~19.50$^{\circ}$), topological insulator (blue, (1 0 20): 2$\theta$~=~66.10$^{\circ}$, offset~=~23.34$^{\circ}$) and capping layer (red, Te (2 1 1): 2$\theta$~=~65.56$^{\circ}$, offset~=~24.78$^{\circ}$) peaks, offset for clarity. The substrate and topological insulator peaks have the expected 3- and 6-fold symmetries, whereas Te (2 1 1) has 6-fold rather than 2-fold symmetry.}
\label{inplane}
\end{figure}

We observe regular growth of the QLs, with homogeneous thickness of $\sim$1~nm for each layer (Figure~\ref{TEM}a and \ref{TEM}b). The Te capping layer is crystalline and appears to grow at an angle of $\sim$7$^{\circ}$ to the underlying quintuple layers. However, detailed HRXRD indicates that this is not the case (see Figure~\ref{inplane} and discussion). While above the first QL we observe good epitaxial growth, the interface between STO and the QLs is relatively irregular, with the first QL being discontinuous and heavy atomic species coexisting with the lighter substrate. The SrTiO$_3$ substrate also displays a change in contrast over a thickness of $\sim$1.5~nm at the substrate surface, where a lower HAADF signal suggests that only light elements are present.

STEM-EDX (energy-dispersive X-ray spectroscopy) elemental maps are reported in Figure~\ref{TEM}c (non-negative matrix factorisation (NMF) of the EDX data is shown in Supplementary Figure S2). The Pt signal originates from the protective layer deposited during focused ion beam (FIB) sample preparation. The QL region contains strong signals from Bi, Sb, V and Te, which are all homogeneous throughout the device thickness. Tellurium extends above the active region into the capping layer. As expected, Sr, Ti and O signals dominate the substrate region. However, Ti and O extend further towards the (Bi,~Sb)$_{2-x}$V$_{x}$Te$_3$ compared to Sr, indicating that the interfacial region that appears darker in the STEM-HAADF images has a lower Sr concentration than the bulk SrTiO$_3$. 

Having confirmed that the films are epitaxial and homogeneously doped, we use reciprocal space mapping of asymmetrical peaks (those whose corresponding atomic planes are not parallel to the substrate) to find the in-plane orientations of the substrate, ferromagnetic topological insulator, and capping layer. Figure \ref{inplane}a-\ref{inplane}c shows reciprocal space maps of a 10~QL, x~=~0.07 sample, where $S_x$ is the reciprocal of the in-plane d-spacing, and $S_z$ is the reciprocal of the out-of-plane d-spacing. The three maps show equivalent areas of reciprocal space at three different azimuthal angles. The maps include the (Bi,~Sb)$_{2-x}$V$_{x}$Te$_3$~(1 0 20) and Te~(2 1 1) peaks (right- and left- hand side of the panels, respectively). Figure \ref{inplane}a also shows (Bi,~Sb)$_{2-x}$V$_{x}$Te$_3$~(1 0 19) directly below (1 0 20). 

Whilst the higher-intensity (1 0 20) peak has the expected six-fold symmetry of a (Bi,~Sb)$_2$Te$_3$-based compound, the appearance of the Te~(2 1 1) peak at all three angles indicates that the capping layer is polycrystalline. Azimuthal angle scans confirm these findings (Figure \ref{inplane}d), and indicate the epitaxial relationship (Bi, Sb)$_{2-x}$V$_{x}$Te$_3$~[0~1~0]~//~SrTiO$_3$~[1~1~$\bar{2}$].
 Note that one of the Te peaks was not captured in this scan due to slight sample misalignment, and that the broader peaks visible on the thin-film scans are due to sample geometry, confirmed by repeating the measurement away from the peaks.
We also observe a Te~(2 1 0) peak at $\Delta\phi \approx$~53$^{\circ}$ from Te~(2 1 1), confirming that the $a$- and $c$-axes of the tellurium cap are in-plane (see Supplementary Figure~S3), and an epitaxial relationship of (Bi,~Sb)$_{2-x}$V$_{x}$Te$_3$~[1~0~0]~//~Te~[$\bar{1}$~2~2]. Since the crystallographic axes are in-plane, we conclude that the $\sim$7.4$^{\circ}$ tilt observed with STEM is not a tilt of the unit cell itself, but rather an alignment of Te atoms revealed by an off-axis cut.

\subsection*{Thickness, doping and substrate dependence of lattice parameters}
\begin{figure}[ht]
\centering
\includegraphics[width=8.5cm]{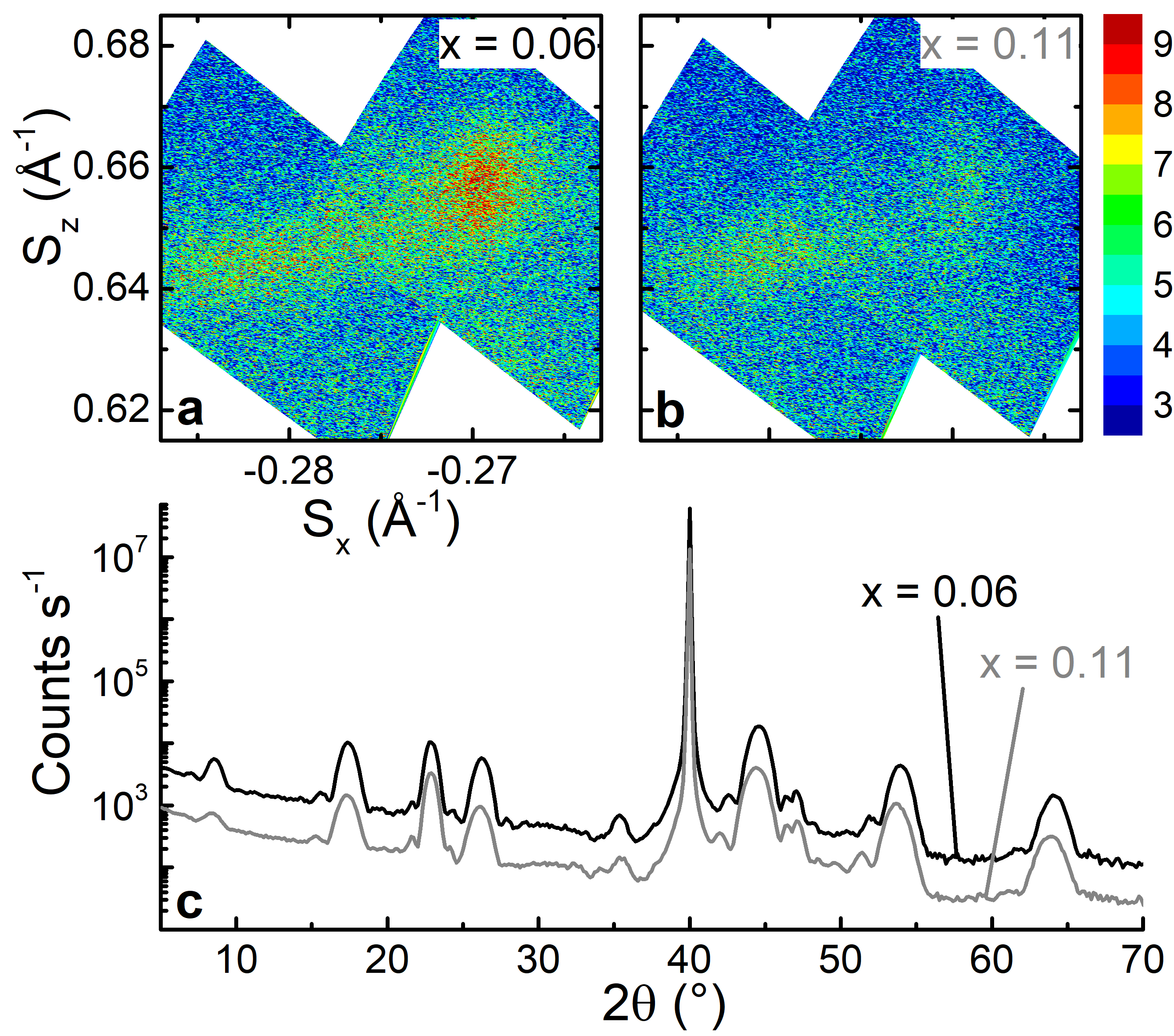}
\caption{Effect of vanadium doping on the unit cell. Reciprocal space maps of the (1 0 20) and Te (2 1 1) (right and left peaks in panels, respectively) in films with \textbf{a} x~=~0.06 and \textbf{b} x~=~0.11, where (Bi,~Sb)$_{2-x}$V$_x$Te$_3$. \textbf{c} 2$\theta$/$\omega$ scans (offset for clarity) of the same films.}
\label{doping}
\end{figure}

\begin{figure}[ht]
\centering
\includegraphics[width=12.75cm]{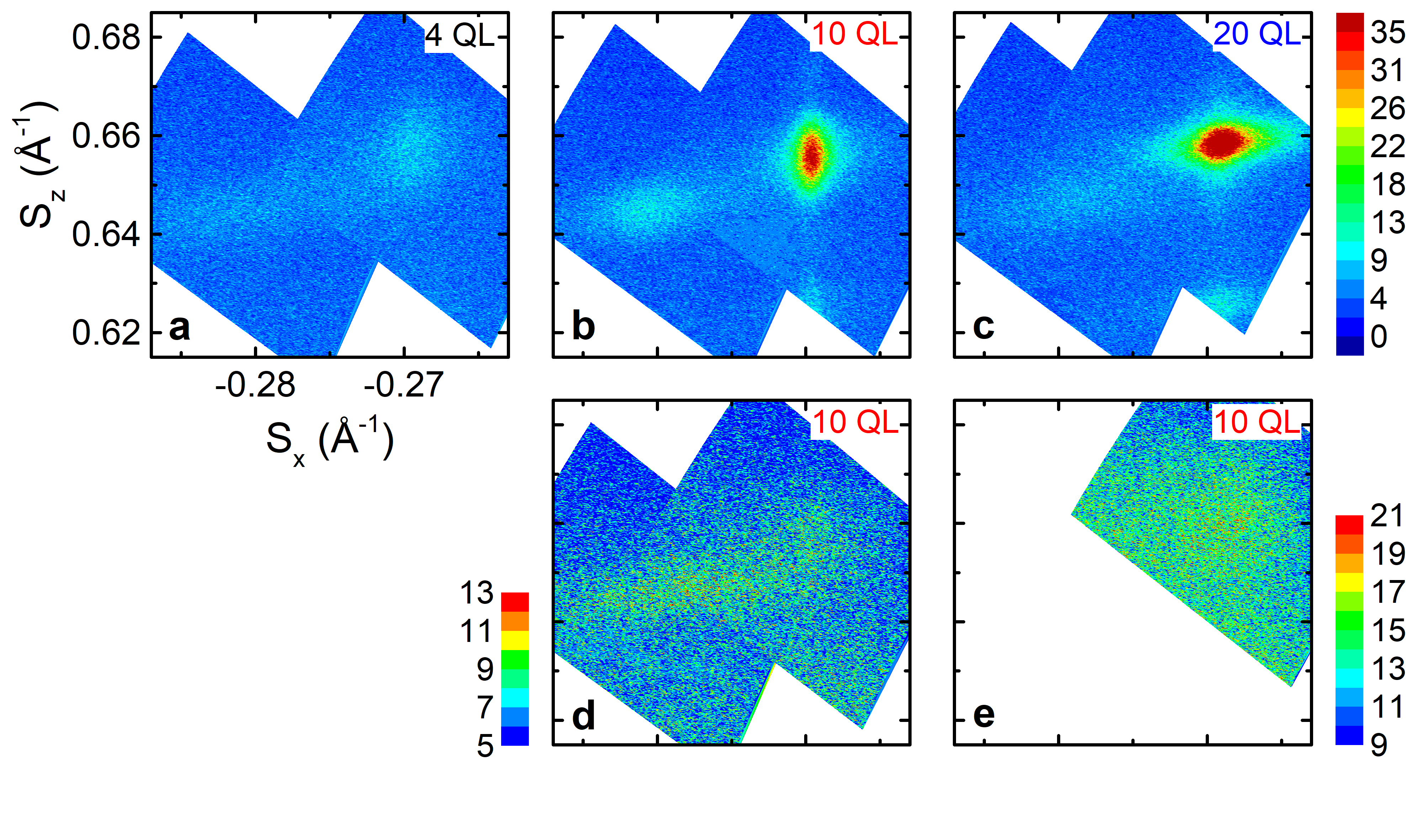}
\caption{Unit cell parameters as a function of thickness and substrate choice. Reciprocal space maps showing the (Bi,~Sb)$_{2-x}$V$_x$Te$_3$ (1 0 20) and Te (2 1 1) peaks, from films grown on SrTiO$_3$ (1~1~1) (\textbf{a-c}: 4, 10 and 20~QL, respectively), Al$_2$O$_3$ (0~0~0~1) (\textbf{d}: 10~QL), and Si (1~1~1) (\textbf{e}: 10~QL). Film peaks measured with respect to SrTiO$_3$~(1~1~2), Al$_3$O$_3$~(0 1 -1 8) or Si~(3~3~1). Panel \textbf{a} shows the same dataset as Figure \ref{doping}\textbf{a}.}
\label{thickness}
\end{figure}

To determine whether the unit cell of (Bi, Sb)$_{2-x}$V$_{x}$Te$_3$ changes with doping level or film thickness, we repeat the reciprocal-space measurements detailed above. A longer counting time is used in order to precisely determine the in- and out-of-plane lattice parameters. Figure \ref{doping} is a comparison of 4~QL films with $x$~=~0.06 and 0.11, and Figure \ref{thickness} shows reciprocal space maps for 4, 10 and 20~QL films on SrTiO$_3$, and 10~QL films on Al$_2$O$_3$~(0~0~0~1) and Si~(1~1~1). Table \ref{table} summarises the calculated lattice parameters, along with previous HRXRD results from the literature for comparison.

Reciprocal-space maps (Figure \ref{doping}a and b, x~=~0.06 and 0.11, respectively) and 2$\theta$/$\omega$ scans (Figure \ref{doping}c, $x$~=~0.11 data (grey) offset for clarity) indicate that the $c$-axis lengthens as the doping is increased. 
From the asymmetrical peaks (Figure \ref{doping}a and \ref{doping}b), $c$~=~30.55~$\pm$~0.42~{\AA} for $x$~=~0.06 and 30.50~$\pm$~0.42~{\AA} for $x$~=~0.11, whereas from the symmetrical peak measurements (Figure\ \ref{doping}c), $c$~=~30.51~$\pm$~0.05~{\AA} and 30.62~$\pm$~0.06~{\AA}. For both doping levels, \emph{c} is still close to the bulk value of 30.60~\AA.

The in-plane lattice parameter is also close to the bulk value ($a$~=~4.30~\AA). For the $x$~=~0.06 and 0.11 films, the $a$-axis parameter is 4.28~$\pm$~0.14~{\AA} or 4.26~$\pm$~0.15~\AA, respectively, and matches to the effective in-plane spacing of the substrate (3.565~\AA) with a 30$^{\circ}$ rotation ($a^{TI}\cos 30^{\circ}$~=~3.71~\AA). The slight discrepancy may be due to the Sr-deficient interfacial layer shown in Figure~\ref{TEM}, which appears to have a larger lattice than the bulk of the substrate.

The in-plane parameter also remains constant as a function of thickness (Figure \ref{thickness}a-\ref{thickness}c): the 10 and 20~QL samples ($x$~=~0.07, Figure \ref{thickness}b and \ref{thickness}c) have $a$~=~4.28~$\pm$~0.16~{\AA} and 4.30~$\pm$~0.16~\AA. Out-of-plane, the thickest (20~QL) film has a smaller $c$, at 30.34~$\pm$~0.41~\AA. The 10 and 4~QL films have similar unit cells, with $c$~=~30.54~$\pm$~0.43~{\AA} and 30.55~$\pm$~0.42~\AA.

Figure~\ref{thickness}b, \ref{thickness}d and \ref{thickness}e shows data taken for 10~QL films grown on SrTiO$_3$ (1 1 1), Al$_2$O$_3$ (0~0~0~1) and Si (1 1 1), respectively. There is very little difference between the in-plane parameters on the three substrates: $a$~=~4.28~$\pm$~0.16, 4.29~$\pm$~0.04 and 4.30~$\pm$~0.20~\AA, respectively. Out-of-plane, the film grown on Si~(1~1~1) appears to have a shorter unit cell, 30.22~$\pm$~0.41~{\AA} as opposed to 30.49~$\pm$~0.32~{\AA} on Al$_2$O$_3$ and 30.54~$\pm$~0.43~{\AA} on SrTiO$_3$. The Te~(2~1~1) peak was not observed for the Si~(1~1~1) sample; based on the relative peak breadths and intensities of symmetric 2$\theta$/$\omega$ scans, the Te capping layer was too thin (2--5~nm) for the peak to be detected (see Supplementary Figure~S1). The peaks measured on Al$_2$O$_3$ (0~0~0~1) and Si (1~1~1) were all less intense and broader than for the equivalent film grown on SrTiO$_3$~(1~1~1), which indicates less well-defined crystallographic orientation in-plane. Whilst the thickness of the films contributes to the breadth and low intensity of the peaks, Scherrer fits to the symmetrical data show that the nominally 10~QL films on SrTiO$_3$ (1 1 1), Al$_2$O$_3$ (0~0~0~1) and Si (1~1~1) are 9.9~$\pm$~1.9~nm, 10.8~$\pm$~0.1~nm and 7.2~$\pm$~0.8~nm, respectively, which does not correlate to the observed difference in intensity.

\section*{Discussion}
\begin{figure}[ht]
\centering
\includegraphics[width=8.5cm]{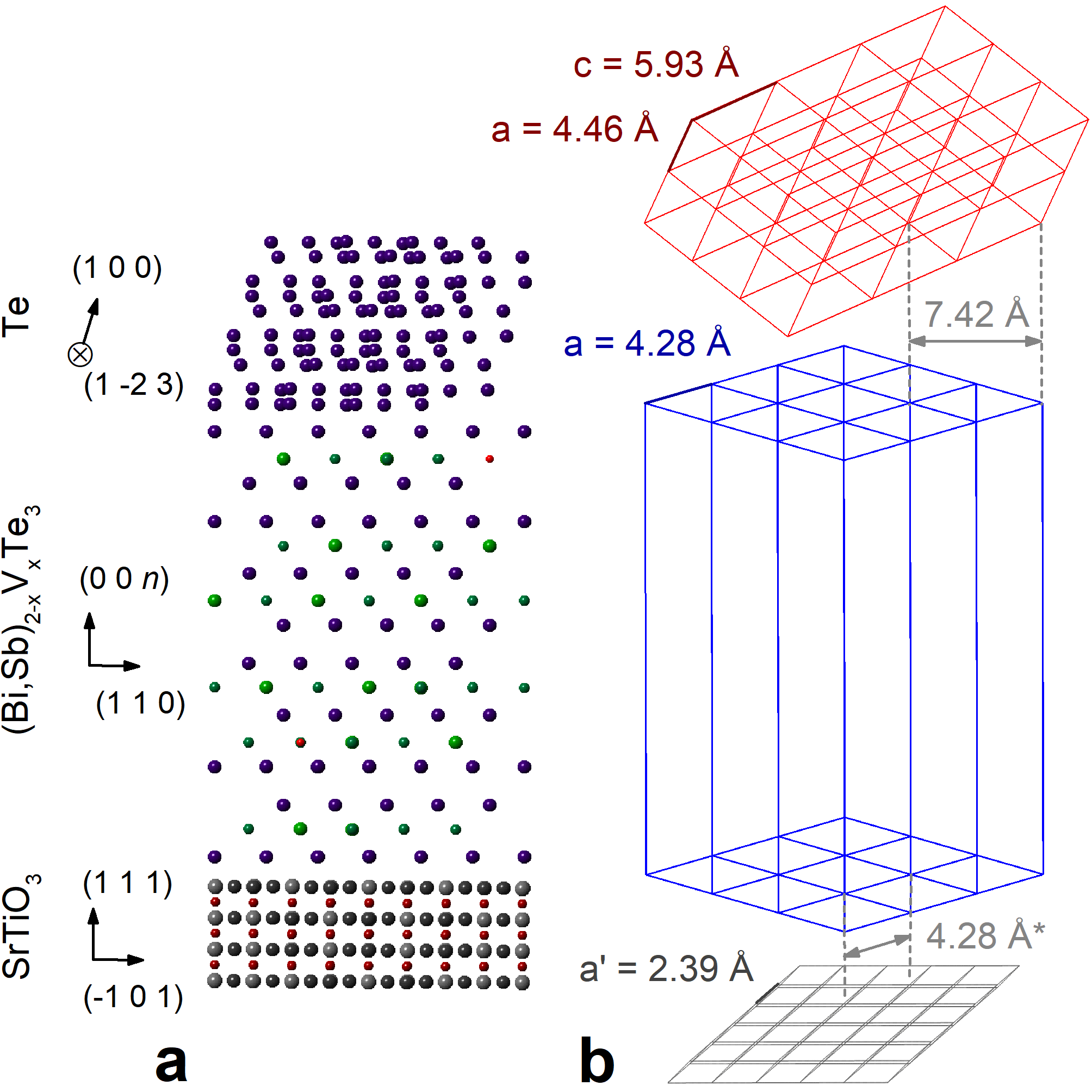}
\caption{Schematic of SrTiO$_3$/(Bi,~Sb)$_{2-x}$V$_x$Te$_3$/Te. \textbf{a} View along the (Bi,~Sb)$_{2-x}$V$_x$Te$_3$ ($\bar{1}$~1~0) direction (Te atoms = dark blue, Sb = dark green, Bi = light green, V = red, Sr = grey, Ti = dark red, O = dark grey). Parameters used are those of the 10~QL film. \textbf{b} Unit cells and experimental lattice parameters of the substrate and films, demonstrating in-plane lattice matching.}
\label{schematic}
\end{figure}

Our results, and those from previous HRXRD studies of ferromagnetic (Bi,~Sb)$_2$Te$_3$, are presented in Table~\ref{table}. Figure \ref{schematic} is a schematic summarizing the results for 10~QL of (Bi,~Sb)$_{2-x}$V$_x$Te$_3$ on SrTiO$_3$, with a Te capping layer.
Surprisingly, measurements of the in-plane lattice constant of these quantum anomalous Hall insulators have only been made for Cr-doped films grown on SrTiO$_3$ (see Table~\ref{table}), even though the QAHE has also been observed in (Bi, Sb)$_{2-x}$Cr$_x$Te$_3$ on InP~(1~1~1) \cite{Checkelsky2012,Checkelsky2014}, Si~(1~1~1) \cite{Grauer2015,Peixoto2016} and GaAs~(1~1~1) \cite{Kou2014,Kou2015}, and grown on Al$_2$O$_3$~(0~0~0~1) \cite{Chang2015a}. Vanadium-doped films have not previously been characterised in-plane.

Although we do not observe significant differences in the (Bi,~Sb)$_{2-x}$V$_x$Te$_3$ unit cell between various doping levels, thicknesses or substrates, there are small trends in \emph{c}, whereas \emph{a} remains unchanged across all the samples measured. The latter indicates that (Bi, Sb)$_{2}$Te$_3$ does not strongly match to the substrate, perhaps due to weak, Van der Waals bonding between quintuple layers and the substrate. Doping with vanadium or chromium slightly lengthens the unit cell along the \emph{c}-axis, though asymmetrical measurements show no difference between our V-doped samples. 
Where data is available for comparison, 20~QL films have a shorter unit cell than 4--10~QL films. This could be due to changing conditions during growth, or perhaps a gradient in the doping after around 10~QL (up to at least 10~QL, V-doping is homogeneous, as shown by EDX). Finally, the unit cells of films grown on Al$_2$O$_3$~(0~0~0~1) and SrTiO$_3$~(1~1~1) are almost identical, but the \emph{c}-axis on Si~(1~1~1) is consistently smaller. This may be because Si~(1~1~1) is not as close a match to the (Bi,~Sb)$_{2-x}$V$_x$Te$_3$ lattice as Al$_2$O$_3$~(0~0~0~1) and SrTiO$_3$~(1~1~1)---rocking curves on the latter samples also show far less disorder (see Supplementary Figure~S1). Although small changes in the unit cell within the parameters used here indicate that strain engineering is not feasible, the lack of biaxial strain in particular is important for growth of doped/undoped heterostructures such as those in Reference~\cite{Mogi2017}.

\begin{table}[ht] \centering \begin{threeparttable}
\begin{tabular}{llclll} \toprule
Substrate & Thickness & Doping & c (\AA, $2\theta$/$\omega$) & c (\AA, RSM) & a (\AA, RSM)\\ \midrule
Bulk & & & & 30.60 & 4.30 \\[10pt]
STO (1~1~1) & 4 QL & V$_{0.06}$ & 30.51 $\pm$ 0.05 & 30.55 $\pm$ 0.42 & 4.28 $\pm$ 0.14 \\
& & V$_{0.11}$ & 30.62 $\pm$ 0.06 & 30.50 $\pm$ 0.42 & 4.26 $\pm$ 0.15 \\[5pt]
& 5 QL\tnote{a} & Cr$_{0.15}$\tnote{*} & 30.62 $\pm$ 0.06 & & 4.28 \\
& & Cr$_{0.15}$ & 30.66 $\pm$ 0.19 & & 4.28 \\[5pt]
& 10 QL & V$_{0.07}$ & 30.54 $\pm$ 0.03 & 30.54 $\pm$ 0.43 & 4.28 $\pm$ 0.16 \\[5pt]
& 20 QL & V$_{0.07}$ & 30.44 $\pm$ 0.01 & 30.34 $\pm$ 0.41 & 4.30 $\pm$ 0.16 \\[10pt]
Al$_2$O$_3$ (0~0~0~1) & 10 QL & V$_{0.07}$ & 30.47 $\pm$ 0.08 & 30.49 $\pm$ 0.32 & 4.29 $\pm$ 0.13 \\[5pt]
& 20 QL\tnote{b} & V$_{0.15}$ & 30.39 & & \\
& & Cr$_{0.16}$ & 30.15 & & \\
& & None & 30.34 & & \\[10pt]
Si (1~1~1) & 10 QL & V$_{0.07}$ & 30.45 $\pm$ 0.07 & 30.22 $\pm$ 0.41 & 4.30 $\pm$ 0.20 \\[10pt]
InP (1~1~1) & 8 QL\tnote{c} & Cr$_{0.22}$ & 30.26 & & \\ \bottomrule
\end{tabular}
\begin{tablenotes}[para,flushleft]
\item[*] Uncapped
\item[a] J. Park \emph{et al.} \cite{Park2015}
\item[b] C.-Z. Chang \emph{et al.} \cite{Chang2015a}
\item[c] J. Checkelsky \emph{et al.} \cite{Checkelsky2014}
\end{tablenotes}
\caption{\label{table} Unit cell parameters of Cr- and V-doped (Bi,~Sb)$_2$Te$_3$ calculated from HRXRD measurements of samples with various substrates, thicknesses and doping levels. Parameters from this study are calculated from three (0~0~\emph{n}) reflections or a single (1~0~20) reflection.}
\end{threeparttable} \end{table}

We also find that the crystal structure of the V-doped films is unchanged compared to Cr-doped (Bi,~Sb)$_2$Te$_3$ \cite{Park2015}, and that the Te capping layer is epitaxial and grows in three equivalent orientations on (Bi,~Sb)$_2$Te$_3$. For all our samples, we observe the epitaxial relationship (Bi,~Sb)$_2$Te$_3$~[1~0~0]~//~Te~[$\bar{1}$~2~2]. The slight out-of-plane disorder in the Te capping layer (observed as a larger FWHM of the rocking curve than (Bi,~Sb)$_{2-x}$V$_x$Te$_3$) is probably due to dislocations at monocrystalline domain boundaries or island formation on the top of the topological insulator during growth (as observed in STEM measurements). 

In summary, we have investigated the crystal structure of the ferromagnetic topological insulator (Bi,~Sb)$_{2-x}$V$_x$Te$_3$ as a function of doping level, thickness and substrate, using high-resolution X-ray diffractometry supported by STEM and EDX. Focusing on the range commonly used in devices, we find that the unit cell is largely unaffected by vanadium doping changes of $\sim$2~at.\%, and remains unchanged over a thickness range of 4--10 quintuple layers (4--10~nm). Substrate choice does not affect the in-plane lattice parameter (\emph{a}), however, out-of-plane the \emph{c}-axis is weakly reduced in films grown on less closely lattice-matched substrates. These results are consistent with previous studies of ferromagnetic topological insulators. We also confirm the previous results of Park \emph{et al.}\cite{Park2015} regarding the Te capping layer growth and orientation, which grows epitaxially in three equivalent orientations on (Bi,~Sb)$_{2}$Te$_3$. Since the in-plane lattice parameter remains constant over this experimentally-relevant range, heterostructures of doped and undoped (Bi,~Sb)$_{2}$Te$_3$ (for example, References \citen{He2016} and \citen{Mogi2017}) are a more promising route to applications than devices which rely on inducing strain to tailor the electronic band structure (as in Reference \citen{Aramberri2016}).

\section*{Methods}
Vanadium-doped bismuth antimony telluride films were grown on SrTiO$_3$~(1~1~1), Al$_2$O$_3$~(0~0~0~1) and Si~(1~1~1) substrates by molecular beam epitaxy, and capped with 10~nm of tellurium (a detailed description is given in Chang et al.~\cite{Chang2015a}).

We performed high-resolution X-ray diffractometry on 4, 10 and 20 quintuple-layer films in a Panalytical Empyrean (Series 2) $\theta$-$\theta$ diffractometer. Optics were optimised for intensity over resolution due to the film thickness. We used a Ge(2 2 0) hybrid monochromator (for Cu$_{\kappa_{\alpha}}$), and a 1/2$^{\circ}$ divergence slit on the incident beam. Diffracted-beam optics comprised either a Xe proportional counter and 1~mm beam tunnel (for symmetrical 2$\theta$/$\omega$ measurements and rocking curves), or a PIXcel$^{3D}$ (Medipix2) detector in 1D frame grab mode (for asymmetrical reciprocal space measurements). Typically, a step time of at least 10s was required to detect the thin-film peaks, whereas the data shown in Figure \ref{thickness} has a step time of 20 minutes. Reciprocal-space maps are measured relative to the STO~(1~1~2), Al$_2$O$_3$~(0~1~-1~8) or Si~(3~3~1) peak.

A cross-sectional sample for TEM analysis was prepared using a dual beam FIB/SEM (FEI Helios Nanolab), from a 10 quintuple-layer film on SrTiO$_3$~(1~1~1). STEM-EDX analysis was carried out in a FEI Osiris, operated at 200~kV, equipped with a set of four EDX detectors in a cross configuration (Super-X by Bruker). Elemental maps were denoised using Principal Component Analysis (PCA) routines integrated in Hyperspy, an open source toolkit for EM data analysis, and the maps shown in Figure \ref{TEM}c are 100x40~px, with a px size 0.5~nm~x~0.5~nm and Gaussian blur 0.5~px for display. The High Resolution STEM images (HRSTEM) were acquired on a probe-corrected FEI Titan with an acceleration voltage of 300~kV.

\footnotesize
\bibliographystyle{clr4}
\bibliography{FMTI_XRD_Refs_v2}

\section*{Acknowledgements}
We would like to thank Dr Ferhat Katmis for useful discussion about the HRXRD data. This work was financially supported by the Leverhulme Trust (RPG-2013-337) and the Royal Society.

\section*{Author contributions statement}
J.M. and J.R. conceived the experiments, C.C. grew the films, and C.R., G.D., M.V., J.S. and M.A. conducted the experiments. C.R. analysed the data, C.R., G.D., J.S. and J.R. wrote the manuscript, and all authors reviewed the manuscript.

\section*{Additional information}
The authors declare no competing financial interests.

\newpage
\section*{Supplementary Information}
\renewcommand{\thefigure}{S\arabic{figure}}
\setcounter{figure}{0}

\begin{figure}[ht]
\centering
\includegraphics[width=11.3cm]{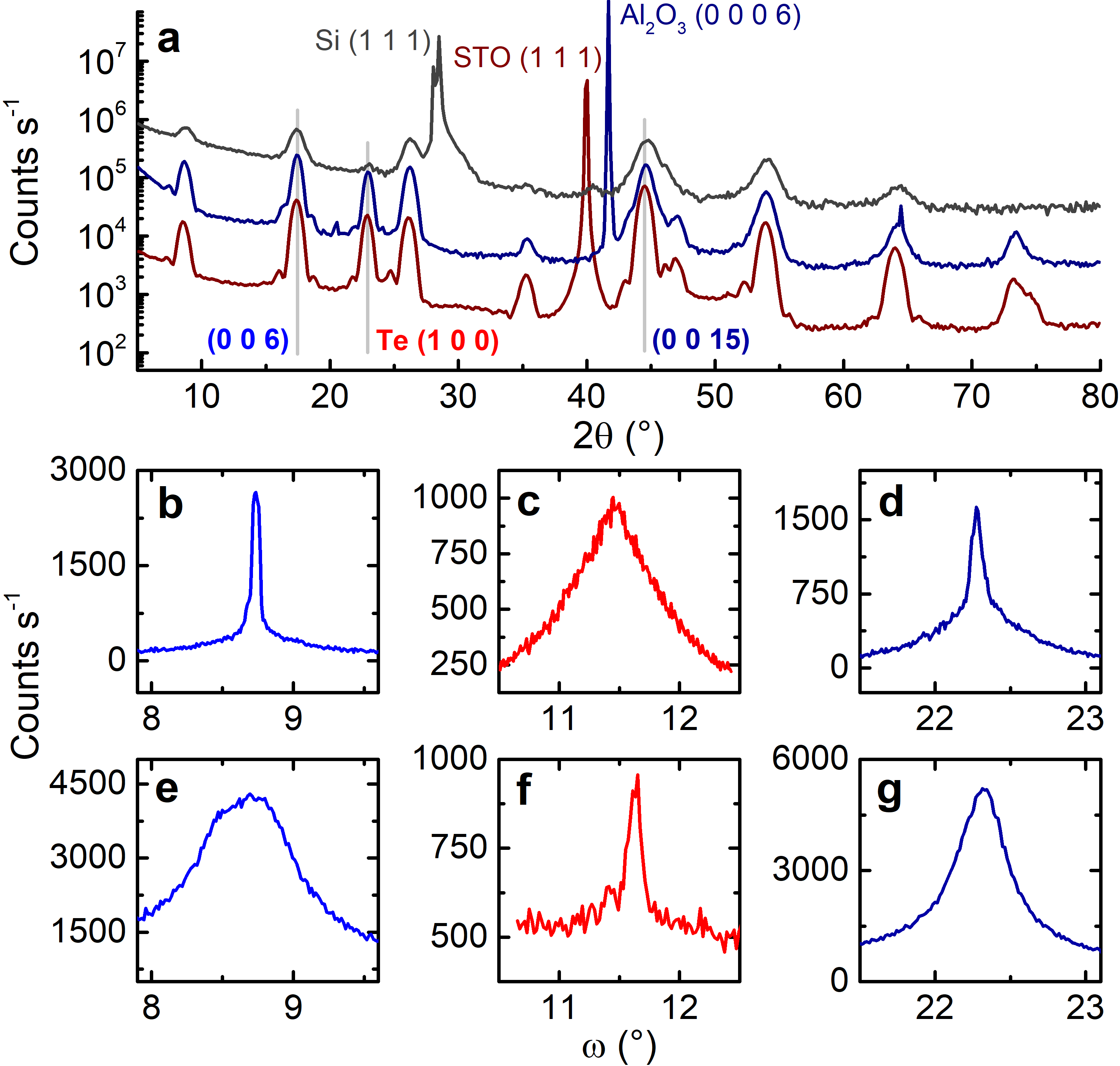}
\caption{HRXRD measurements of 10~QL (Bi,~Sb)$_{2-x}$V$_x$Te$_3$ films grown on SrTiO$_3$~(1~1~1), Al$_2$O$_3$~(0~0~0~1) and Si~(1~1~1). \textbf{a} 2$\theta$/$\omega$ scans of films grown on SrTiO$_3$~(1~1~1) (dark red), Al$_2$O$_3$~(0~0~0~1) (dark blue) and Si~(1~1~1) (grey). The much lower intensity of the Te~(1~0~0) peak on Si~(1~1~1) indicates a capping layer thickness below 5~nm. \textbf{b-d} Rocking curves from the Al$_2$O$_3$~(0~0~0~1) sample, taken on the (Bi,~Sb)$_{2-x}$V$_x$Te$_3$ (0 0 6), Te (1 0 0) and (Bi,~Sb)$_{2-x}$V$_x$Te$_3$ (0 0 15) peaks, respectively. The rocking curve widths are very similar to those of the SrTiO$_3$~(1 1 1) sample. \textbf{e-g} Equivalent rocking curves from the Si~(1~1~1) sample, with a larger width indicating more disordered growth of the (Bi,~Sb)$_{2-x}$V$_x$Te$_3$ on this substrate.}
\end{figure}

\begin{figure}[ht]
\centering
\includegraphics[width=\linewidth]{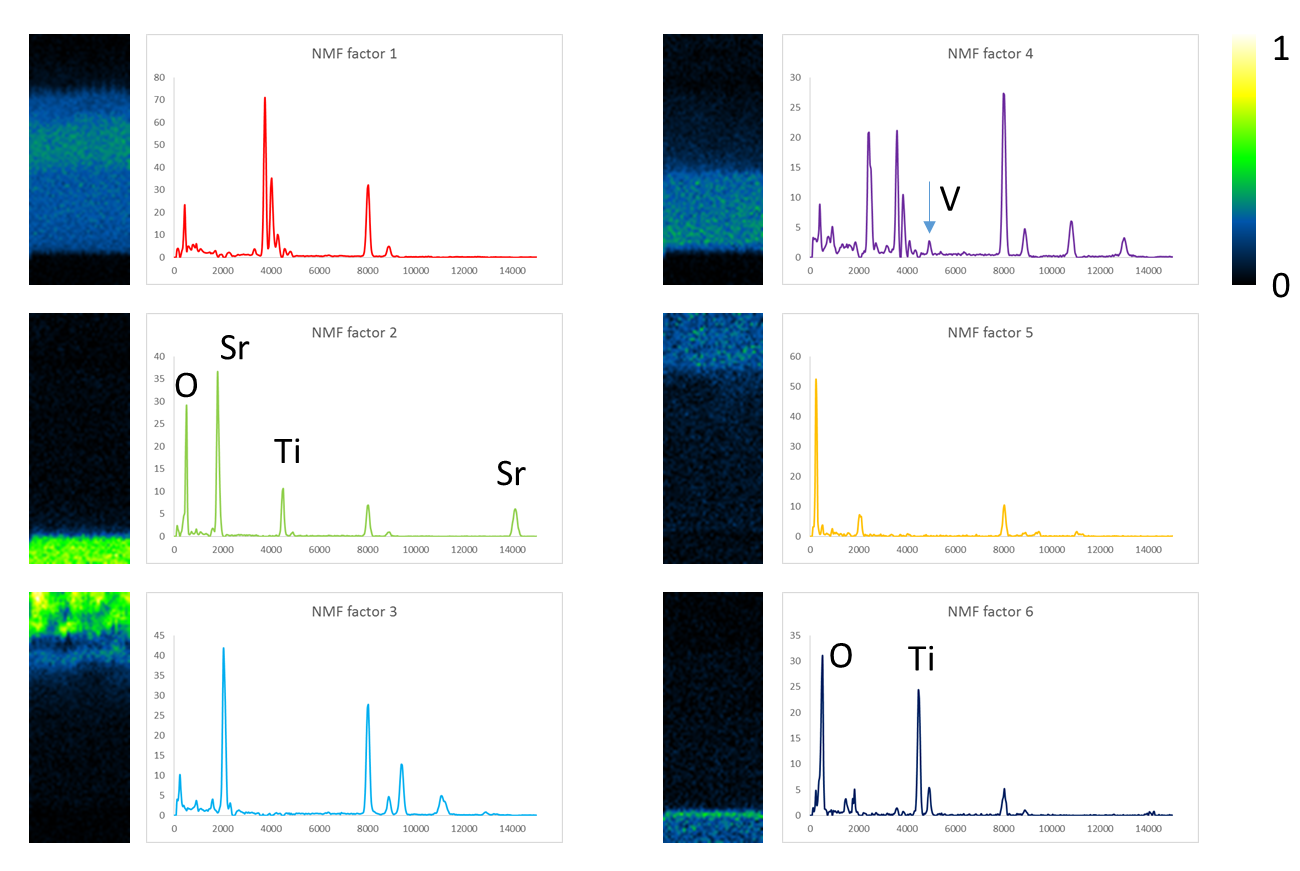}
\caption{Non-negative matrix factorisation (NMF) of the EDX dataset, showing layer composition. Factor 4 (upper right) contains the QL region of the film, and has features corresponding to all the expected elements (Bi, Sb, Te, V). Factor 6 (lower right) represents a combination of Ti and O (with very little Sr) and is concentrated at the SrTiO$_3$/(Bi,~Sb)$_{2-x}$V$_x$Te$_3$ interface.}
\end{figure}

\begin{figure}[ht]
\centering
\includegraphics[width=11.3cm]{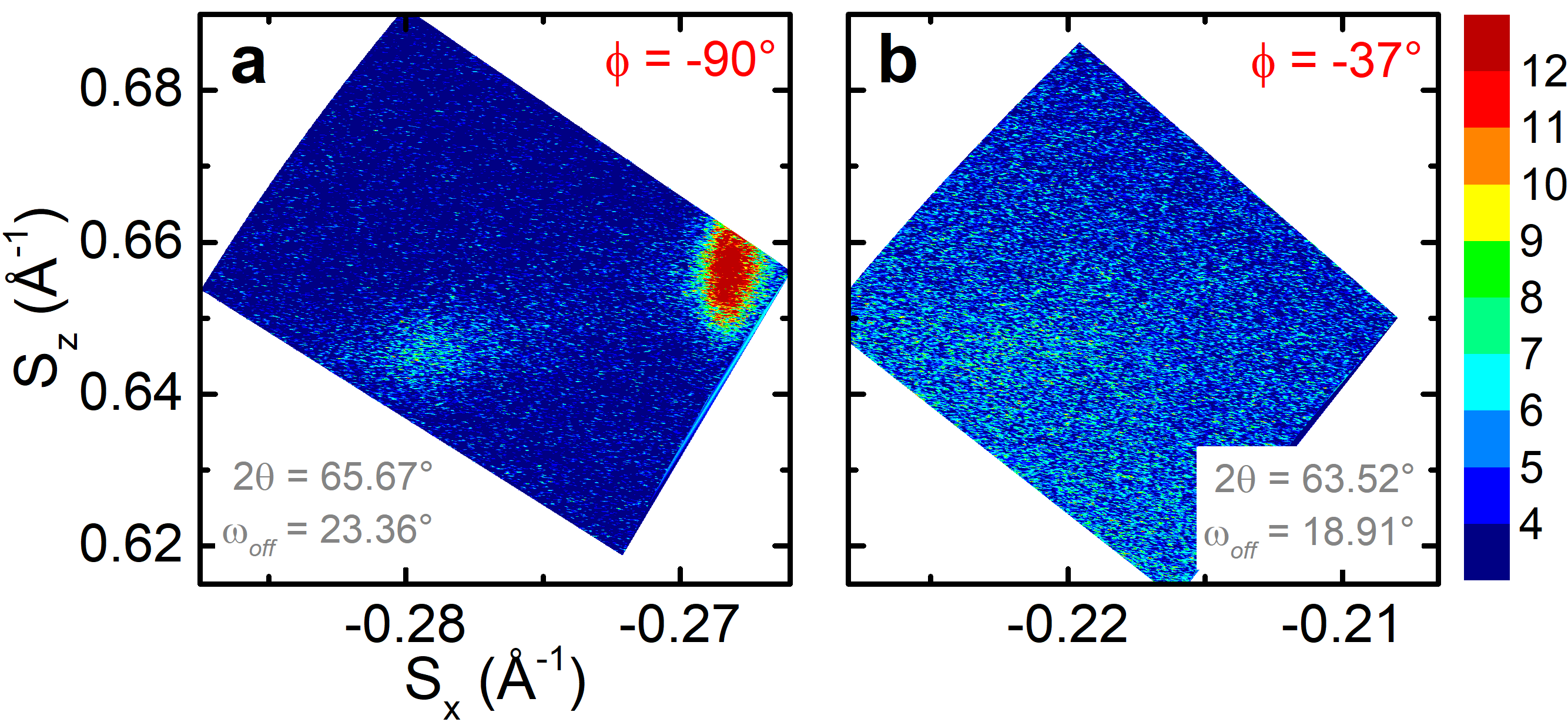}
\caption{HRXRD confirming in-plane orientation of Te capping layer. \textbf{a} Reciprocal-space map showing Te~(2~1~1) and (Bi,~Sb)$_{2-x}$V$_x$Te$_3$ peaks at $\phi$~=~-90$^{\circ}$. \textbf{b} Te~(2~1~0) peak measured at $\phi$~=~-37$^{\circ}$. $\Delta\phi \approx 53^{\circ}$, matching the calculated difference in $\phi$ (53.05$^{\circ}$) between the two peaks when Te~(1~0~0) is normal to the substrate and confirming that the \emph{a}- and \emph{c}-axes are in the plane of the substrate.}
\end{figure}

\end{document}